\documentclass[a4paper,fleqn]{cas-dc}
\usepackage[square,numbers]{natbib}
\usepackage{amsmath,mathrsfs,mathtools,mathdots,amssymb,commath,bm}
\usepackage{multirow,makecell,booktabs,threeparttable}
\usepackage{subfigure}
\usepackage{nomencl}
\usepackage{pifont}
\usepackage{ulem}

\makeatletter

\newcommand{\Rmnum}[1]{\expandafter\@slowromancap\romannumeral #1@}
\makeatother

\def\tsc#1{\csdef{#1}{\textsc{\lowercase{#1}}\xspace}}
\tsc{WGM}
\tsc{QE}
\tsc{EP}
\tsc{PMS}
\tsc{BEC}
\tsc{DE}

\begin{document}
\begin{sloppypar}

\let\WriteBookmarks\relax
\def\floatpagepagefraction{1}
\def\textpagefraction{.001}

\shorttitle{Numerical study of the splashing wave induced by a seaplane using mesh-based and particle-based methods}
\shortauthors{Xu et~al.}

\title [mode = title]{Numerical study of the splashing wave induced by a seaplane using mesh-based and particle-based methods}
\address[1]{School of Ocean Engineering and Technology, Sun Yat-sen University, Zhuhai 519082, China}
\address[2]{CNR-INM. Institute of Marine Engineering, Rome, 00128, Italy}
 \address[3]{AVIC GA Huanan Aircraft Industry Co., Ltd, Zhuhai 519080, China}


\author[1]{Yang Xu}
\ead{xuyang73@mail2.sysu.edu.cn}
\credit{Formal analysis, Validation, Investigation, Visualization, Writing - Original draft preparation}

\author[1]{Peng-Nan Sun}[orcid=0000-0001-8886-6260]
\ead{sunpn@mail.sysu.edu.cn}
\credit{Conceptualization, Methodology, Software, Resources, Data Curation, Supervision, Funding acquisition, Writing - Review \& Editing}
\cormark[1]
\cortext[cor1]{Corresponding author}

\author[1]{Xiao-Ting Huang}
\ead{xiaotinghuanghxt@163.com}
\credit{Visualization, Writing - Review \& Editing}

\author[2]{Salvatore Marrone}
\ead{salvatore.marrone@cnr.it}
\credit{Project administration, Writing - Review \& Editing}


\author[3]{Lei-Ming Geng}
\ead{salvatore.marrone@cnr.it}
\credit{Project administration, Writing - Review \& Editing}

\begin{abstract}
  In recent years, forest fires and maritime accidents have occurred frequently, which have had a bad impact on human production and life. Thus, the development of seaplanes is an increasingly urgent demand. It is important to study the taxiing process of seaplanes for the development of seaplanes, which is a strong nonlinear fluid-structure interaction problem.  In this paper, the Smoothed Particle Hydrodynamics (SPH) method based on the Lagrangian framework is utilized to simulate the taxiing process of seaplanes, and the SPH results are compared with those of the Finite Volume Method (FVM) based on the Eulerian method. The results show that the SPH method can not only give the same accuracy as the FVM but also have a strong ability to capture the splashing waves in the taxiing process, which is quite meaningful for the subsequent study of the effect of a splash on other parts of the seaplane.
\end{abstract}


%


\maketitle

\section{Introduction}
Seaplanes have shown great application potential in forest fire fighting and maritime rescue. The ability to take off and land on the water surface is the most typical feature of a seaplane, which is different from a roadbed aircraft. Takeoff and landing on the water are accompanied by taxiing motion on the water \citep{canamar_leyva_seaplane_2012}. The water surface environment is different from the land runway. Many factors will affect the performance of the seaplane, which may cause the seaplane to be buried or move like a dolphin. It is very important to study the hydrodynamic performance of the seaplane during the taxiing process for the whole design, and it is also of great significance to the subsequent design of large firefighting/water rescue seaplanes.

\begin{figure*}
  \centering
  \includegraphics[width=0.980\linewidth]{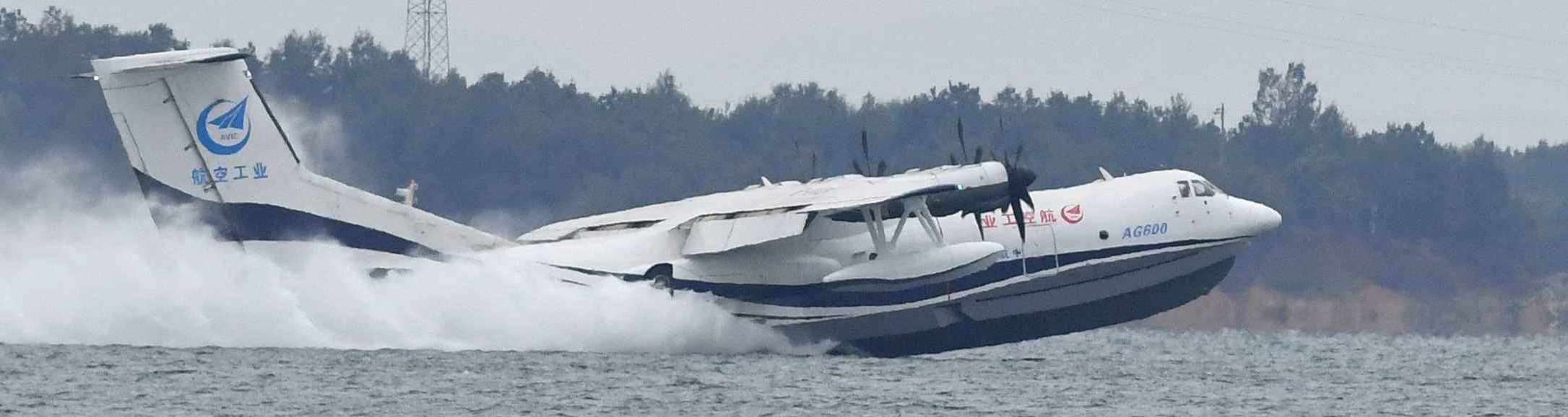}
  \caption{The seaphane of "AG600". Photo from AVIC GA Huanan Aircraft Industry Co., Ltd } 
  \label{fig:AG600-network}
\end{figure*}

At present, the research on the hydrodynamic force of the seaplane during taking-off and landing is mainly divided into model experiments and empirical formulas \citep{von_karman_impact_1929,zheng_experiment_2019}.
With the help of some high-precision experimental equipment, physical experiments can capture the splash generated during the taxiing of seaplanes and provide a more practical assessment. However, the scale effect of the model experimental tank and the high experimental cost are unavoidable topics.
Recently, due to a huge increase in computing power, a variety of numerical methods have been proposed and applied to the study of the hydrodynamics of water entries \citep{zhong_sph_2023,huang_water_2022,liu_investigation_2023,sun_inclusion_2023} and taxiing process of seaplanes \citep{xiao2017development,duan_numerical_2019,zheng_numerical_2021,xiao_effect_2021,chen_numerical_2022,wen_influence_2022}.
Among them, the representative is the Finite Volume Method (FVM).
Based on the commercial software FLUENT, \citet{zhaoyunke2020} simulated the entire process of seaplane landing on the water surface and captured the change of free surface.
In \citet{zheng_numerical_2021}, the numerical simulations of the taxiing process of seaplanes before take-off on the calm water at different speeds were carried out based on STAR-CCM+, with overset mesh technology and VOF method, which is consistent with the experimental results.

There are complex fluid movements such as free surface splashing, overturning, and crushing during the taxiing process of seaplanes. The violent splash of the free surface will have a certain impact on the belly and wings of seaplanes, which may cause certain harm to the safety performance of seaplanes. 
As a class of mesh-free methods based on the Lagrangian view, Smoothed Particle Hydrodynamics (SPH) is very suitable for dealing with free liquid surface tracking \citep{zhang_smoothed_2017,cheng_towards_2019,di_mascio_sphfv_2021,xiao_effect_2021}, water-gas interface \citep{sun_accurate_2018}, and fluid-structure interface \citep{Khayyer2018Enhanced,marrone_high-speed_2018,ju_study_2023}.
\citet{cheng_towards_2019} adopted the SPH model to tackle the ditching problems of a ground-effect wing ship and some useful conclusions were drawn. 
\citet{xiao_effect_2021} predicted the hydrodynamic behavior of ditching events in terms of hydrodynamic forces, pressure loads, and splashing jet topology by means of an SPH model and analyzed the impact of different initial ditching angles.

As mentioned before, the splashing phenomenon during high-speed taxiing of seaplanes is a structure-fluid interaction problem involving solid-liquid-gas, in which gas also occupies a certain influence.
Many scholars have studied the behavior of gas, especially the bubble in the fluid \citep{zhang_unified_2023}. 
 \citet{zhang_influences_2013} accurately simulated the behaviors of the entrained bubble in complex flows and discovered the influence mechanism of fluid properties on the bubble dynamics, which provided good guidance for the research of complex water-gas multiphase flows.
At present, the research focus of this paper is on seaplanes and water, and the influence of air will be considered in subsequent research.

In this work, based on the SPH method and the FVM method, the process of seaplane taxiing on the still water surface is simulated. The feasibility of the SPH method to simulate the taxiing process of the seaplane is analyzed and its ability to capture free surface splash is emphasized.

\section{Numerical methods}
\subsection{Governing equation } 
The solution to fluid dynamics problems is mainly based on a system of partial differential equations for the fields of variables such as density, velocity, and energy. In this paper, the system of SPH theory under the weakly compressible assumption will be described in terms of the basic governing equations. The governing equations \citep{liu_smoothed_2003} for a positive pressure fluid based on the weakly compressible assumption are as follows:
\begin{equation}\label{eq:governing-equation}
  \begin{dcases}
    \frac{\Dif \rho}{\Dif r}=-\rho \nabla \cdot \mathbf{v}\,, \\
    \frac{\Dif u}{\Dif t}=\frac{\nabla \cdot \boldsymbol{T}}{\rho}+\boldsymbol{g}\,, \\
    \frac{\Dif r}{\Dif t}=\mathbf{v}\,, \\
    p=c_{0}{ }^{2}\left(\rho-\rho_{0}\right)
    \end{dcases}
\end{equation}
where $\frac{\Dif}{\Dif t}$ is the material derivative concerning time or the Lagrangian derivative.
$\rho$, $\mathbf{v}$, $\bm{r}$ denote the density, velocity, and position vector of the fluid, respectively. $\boldsymbol{g}$ is the acceleration of gravity. $\boldsymbol{T}$ is the stress tensor of a Newtonian fluid. 
To solve the above governing equations in the SPH framework, these equations need to be discretized by particles, which are described later. 
\subsection{SPH discretization of governing equations}
In the SPH framework, the computational domain is discretized into particles carrying physical properties such as density, pressure, velocity, etc \citep{Liu2010Smoothed}.
The SPH equation is constructed in two steps, the first being the continuous integral representation, also known as the kernel approximation.
The continues integral representation of a field function $f(\boldsymbol{r})$  \citep{liu_smoothed_2003} is:
\begin{equation}
  \begin{dcases}
    f(\boldsymbol{r}) \approx\langle f(\boldsymbol{r})\rangle=\int_{\Omega} f\left(\boldsymbol{r}^{\prime}\right) W\left(\boldsymbol{r}-\boldsymbol{r}^{\prime}, h\right) d \boldsymbol{r}^{\prime} \\
    \nabla_{r} f(\boldsymbol{r}) \approx\left\langle\nabla_{r} f(\boldsymbol{r})\right\rangle = \int_{\Omega} f\left(\boldsymbol{r}^{\prime}\right) \nabla_{r} W\left(\boldsymbol{r}-\boldsymbol{r}^{\prime}, h\right) d \boldsymbol{r}^{\prime}
    \end{dcases}
\end{equation}
where $W$ is the kernel function.
Since the Wendland C2 kernel function has many excellent features, it is selected as the kernel function here.
 $\nabla_{{r}}$ represents the gradient of relative position $\bm{r}$, $h$ smoothing length.
The next thing to do is particle approximation \citep{liu_smoothed_2003}, turning the continuity integral into a discrete sum of particles as follows:
\begin{equation}\label{eq:continuous-internal}
  \left\{\begin{array}{l}
    f\left(\boldsymbol{r}_{i}\right)=\sum_{j} f\left(\boldsymbol{r}_{j}\right) W\left(\boldsymbol{r}_{i}-\boldsymbol{r}_{j}, h\right) V_{j} \\
    \nabla_{i} f\left(\boldsymbol{r}_{i}\right)=\sum_{j}\left[f\left(\boldsymbol{r}_{j}\right)+f\left(\boldsymbol{r}_{j}\right)\right] \nabla W\left(\boldsymbol{r}_{i}-\boldsymbol{r}_{j}, h\right) V_{j} \\
    \nabla_{i} \cdot \boldsymbol{f}\left(\boldsymbol{r}_{i}\right)=\sum_{j}\left[\boldsymbol{f}\left(\boldsymbol{r}_{j}\right)-\boldsymbol{f}\left(\boldsymbol{r}_{j}\right)\right] \cdot \nabla_{i} W\left(\boldsymbol{r}_{i}-\boldsymbol{r}_{j}, h\right) V_{j}
    \end{array}\right.  
\end{equation}
In the above equation, the subscripts $i$, $j$ are the particle numbers; $j$ represents any other particle number in the tightly-branched domain of the particle $i$. 
$V_j$ represents the volume of the particle $j$ and can be obtained from its mass and density.
The Eq.\eqref{eq:continuous-internal} shows the particle approximation of the function and its gradient and divergence.

\subsection{$\delta$-SPH model}
The governing equations \citep{marrone_-sph_2011} are derived as follows:
\begin{equation}\label{eq:delta-SPH}
  \begin{dcases}
    \frac{\Dif\rho_{i}}{\Dif t}
    =-\rho_{i}\sum_{j}\left(\mathbf{v}_{j}-\mathbf{v}_{i}\right)\cdot\nabla_{i}W_{ij}V_{j}
    +\mathcal{T}_{i}^{\left(D\right)}\,, \\
    \frac{\Dif\mathbf{v}_{i}}{\Dif t}
    = -\frac{1}{\rho_{i}}\sum_{j}\left(p_{i}+p_{j}\right)\nabla_{i}W_{ij}V_{j}+\boldsymbol{g}
    +\mathcal{T}_{i}^{\left(V\right)}\,, \\
    \frac{\Dif\mathbf{r}_{i}}{\Dif t}=\mathbf{v}_{i}\,,
    \quad V_{i}\left(t\right)=\frac{m_{i}}{\rho_{i}\left(t\right)}\,,\\
    p_i = c_0\left(\rho_i - \rho_0\right)
  \end{dcases}
\end{equation}
where $p$ is the pressure. 
In this study, the $\delta$-SPH model \citep{sun_numerical_2015,sun_accurate_2021} is adopted, which is used widely in the SPH community. Its accuracy and robustness in simulating fluid-solid coupling problems, especially when it involves large deformation of the water surface, meet the calculation requirements 
The density diffusive term, introduced in the continuity equation, as the most critical superiority of the $\delta$-SPH model can remove the pressure noise efficiently, which is written as $\mathcal{T}_{i}^{\left(D\right)}$ in Eq. \eqref{eq:delta-SPH}.
$\mathcal{T}_{i}^{\left(D\right)}$ is given in detail as:
 \begin{equation}
 \begin{dcases}
    \mathcal{T}_{i}^{(D)}=\delta c_{0} h_{i} \sum_{j} \pi_{i j} \cdot \nabla_{i} W_{i j} V_{j}, \\
    \pi_{i j}=2\left(\rho_{j}-\rho_{i}\right) \frac{\mathbf{r}_{j}-\mathbf{r}_{i}}{\left\|\mathbf{r}_{j}-\mathbf{r}_{i}\right\|^{2}}-\left[\langle\nabla \rho\rangle_{i}^{L}+\langle\nabla \rho\rangle_{j}^{L}\right], \\
    \end{dcases}
  \end{equation}
where the coefficient $\delta$ is usually adopted as 0.1 and $\langle\nabla\rho\rangle^{L}$ represents the renormalized density gradient \citep{antuono_free-surface_2010}.
Moreover,  an artificial viscosity term $\mathcal{T}_{i}^{(V)}$ (\cite{marrone_-sph_2011}) is taken into the momentum equation to guarantee numerical stability, which is given as
  \begin{equation}
    \begin{dcases}
      \mathcal{T}_{i}^{\left(V\right)}=\frac{\alpha c_{0}h_{i}\rho_{0}}{\rho_{i}}\sum_{j}\Pi_{ij}\nabla_{i}W_{ij}V_{j}\,, \\
      \Pi_{ij}=\frac{\left(\mathbf{v}_{j}-\mathbf{v}_{i}\right)\cdot\left(\mathbf{r}_{j}-\mathbf{r}_{i}\right)}{\left\Vert\mathbf{r}_{j}-\mathbf{r}_{i}\right\Vert^{2}+0.01h^{2}}\,,
    \end{dcases}
  \end{equation}
$\alpha$ is set to $0.02$.
In order to solve the above governing equation, it is also necessary to decouple density $\rho$ and pressure ${p}$, which is implemented here using the linearized Tait \textit{Equation of State} (EoS).
It means the pressure directly depends on the reference density seeing the last line in Eq.\eqref{eq:delta-SPH}
where $c_{0}$ represents the initial artificial speed of sound and $\rho_{0}$ represents the initial fluid density.
In the WCSPH framework, the value of $c_{0}$, according to the limitation of Mach number \citep{antuono_energy_2015}, should guarantee that 
\begin{equation}
  c_{0}\ge10\max\left(U_{\text{max}},\sqrt{\frac{P_{\text{max}}}{\rho_{0}}}\right)\,,
\end{equation}
where $U_{\text{max}}$ and $P_{\text{max}}$ stand for the maximum anticipated fluid velocity and pressure during the SPH simulation, respectively.

\begin{figure}
  \centering
  \includegraphics[width=0.950\linewidth]{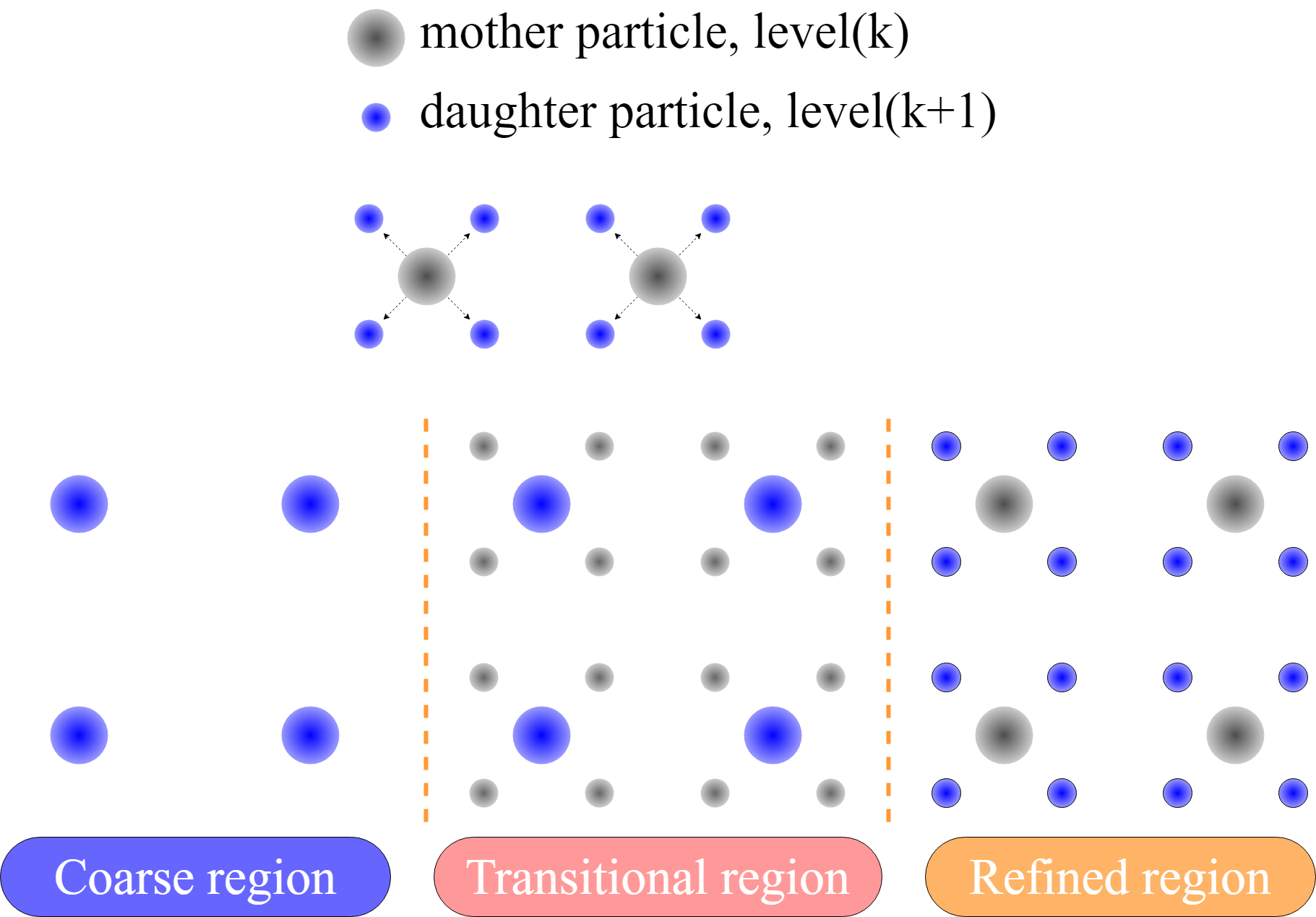}
  \caption{Schematic diagram of adaptive particle refinement. } \label{fig:APR-diagram}
\end{figure} 

\begin{figure*}
  \centering
  \includegraphics[width=0.980\linewidth]{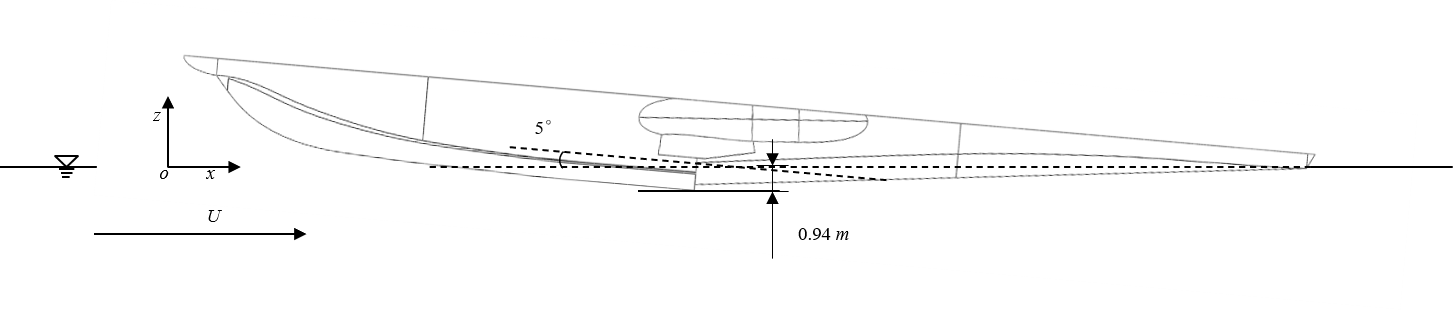}
  \caption{ The initial state of the seaplane ( with a 5-degree angle to the water surface ). } \label{fig:Seaphane-initial}
\end{figure*} 

\begin{figure*}
  \centering
  \includegraphics[width=0.980\linewidth]{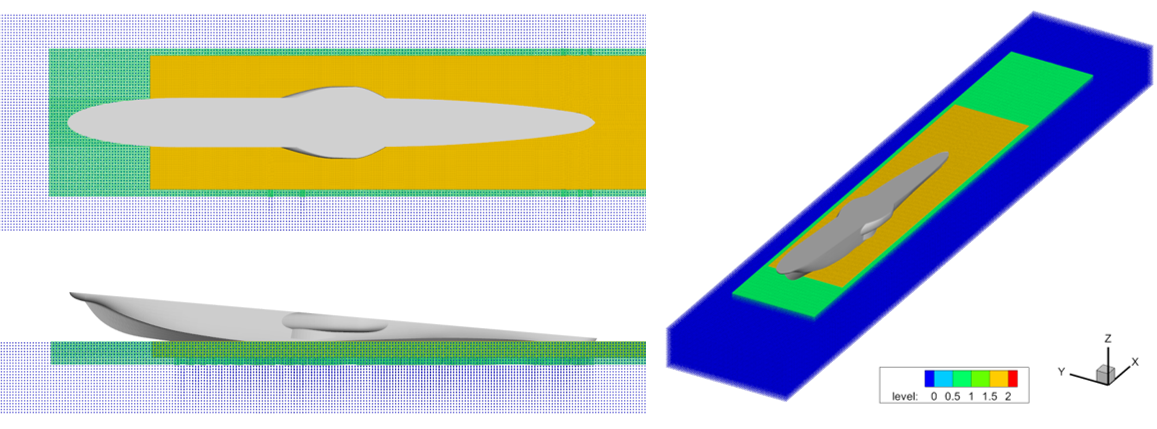}
  \caption{Configuration of seaplane taxiing with high speed simulated with SPH methods combined with APR. On the left, from top to bottom, the top view and side view are given respectively, and the right is a full view of the simulation tank.   } \label{fig:APR-Seaphane-SPH}
\end{figure*} 

\begin{figure*}
  \centering
  \includegraphics[width=0.980\linewidth]{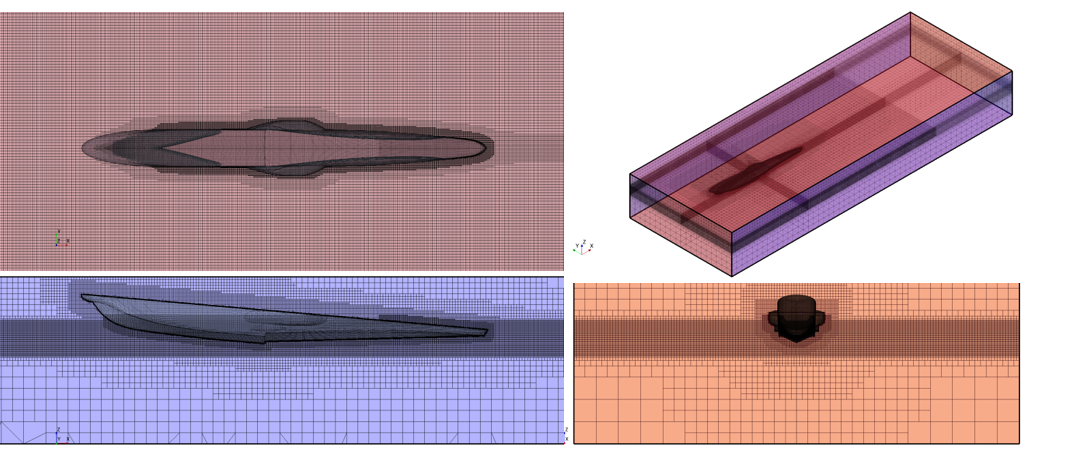}
  \caption{Configuration of seaplane taxiing with high speed simulated with FVM methods with refined mesh. On the left, from top to bottom, the top view and side view are given respectively, while the right is a full view and front view of the simulation tank. } \label{fig:APR-Seaphane-FVM}
\end{figure*}

\subsection{Multi-resolution technology}
In order to improve the computational accuracy and reduce the total computational cost, the technique of adaptive particle refinement (hereafter is shorten to APR) \citep{chiron_analysis_2018,sun_multi-resolution_2018,lyu_3d_2022} is imposed in this paper.  This technique was inspired by the adaptive mesh refinement (AMR) developed in mesh-based models. 
Particles of different resolutions are defined as different sets of particles, denoted by ${level}\left(k\right)$ where $k$ represents the level of particle resolution, equaling the time of particle splitting. 
For example, the particles of the bottom layer belong to ${level}\left(0\right)$. The particle splitting occurs when the particles enter the refinement region. In the three-dimensional problem, a parent particle (${level}\left(k\right)$) splits into eight sub-particles (the splitting process is shown in ). During the splitting process, the initial particles are still retained, generating a new set of particles set as $ {level}\left(k + 1\right)$. The particles of the layer ${level}\left(k\right)$ are called parent particles, while the particles of the layer ${level}\left(k + 1\right)$ are child particles (shown in Fig.\ref{fig:APR-diagram}).
Importantly, during the SPH simulation, the parent particles are turned off as inactive, i.e. they no longer participate in the SPH simulation, but still update particle information and follow the flow field motion by interpolating the data from the daughter particles within the support domain.
The daughter particle formed by the splitting is active and begins to participate in the flow field calculation, also named an SPH particle. Arbitrary physical  represents $\phi$ (e.g. pressure, velocity, density, etc.) for inactive particles can be interpolated from the surrounding SPH particles using the Shepard interpolation method \citep{shepard_two-dimensional_1968} with the following equation:
\begin{equation}
  \phi_{i}=\sum_{j \in S P H} \phi_{j} W_{i j} / \sum_{j \in S P H} W_{i j}
\end{equation}

It is also specified that particles in every layer only interact with particles in the same layer, which means that at the edge of ${level}\left(k + 1\right)$, a transition region needs to be extended (as shown in Fig.\ref{fig:APR-diagram}) with a width greater than the radius of the kernel function in layer ${level}\left(k + 1\right)$, for the sake of preventing errors due to truncation of the kernel function. 
In the transition region, the particle activity is opposite to its interior.
In particular, when the inactive parent particles are out of the particle refinement region, they are reactivated and become SPH particles and then participate in the SPH calculation normally, while the refined daughter particles flow out of the refinement region and its transition region, they are removed. 
The principle and implementation of multi-level resolution are introduced in detail above, through which the multi-level resolution model can be implemented in the SPH model to improve the local calculation accuracy.

\begin{figure*}
  \centering
  \subfigure[Drag force of simulation with speed 30 m/s]{
    \includegraphics[width=0.48\linewidth]{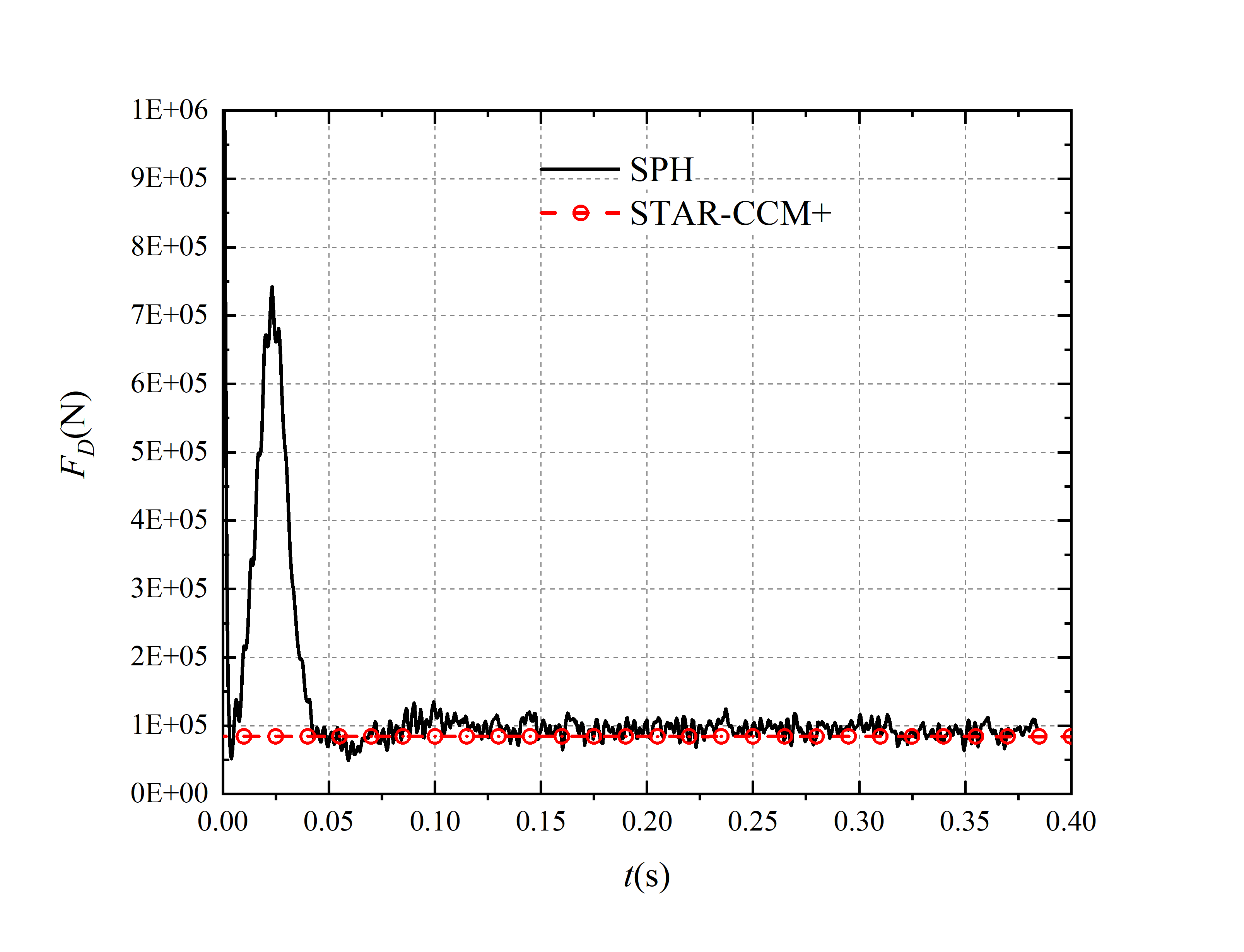}\label{fig:30mps-Drag}
  }
  \subfigure[Lift force of slimulation with speed 30 m/s]{
    \includegraphics[width=0.48\linewidth]{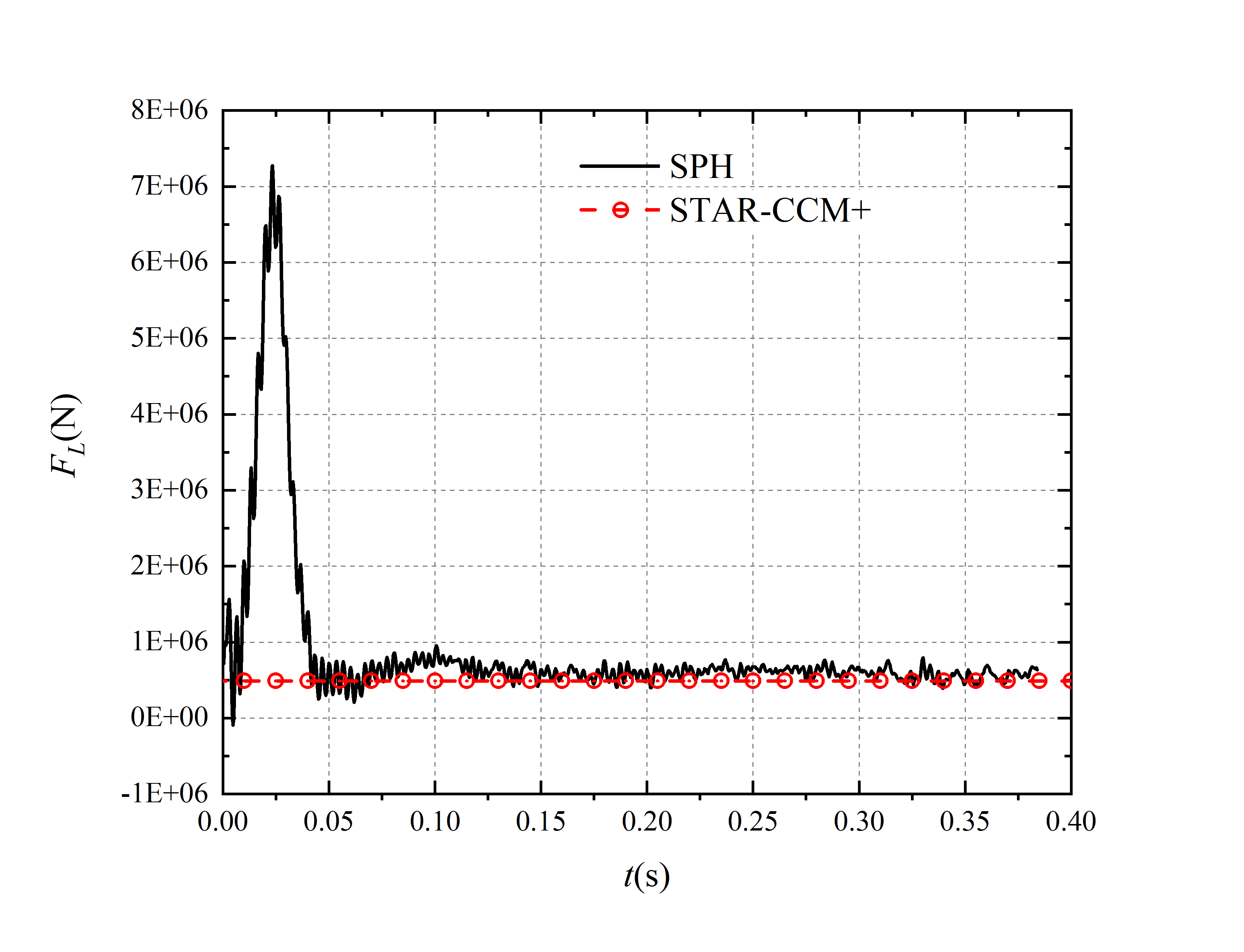}\label{fig:30mps-Lift}
  }
  \caption{The SPH results of drag and lift force validated with FVM results with a speed of 30 m/s.}
  \label{fig:30mps}
\end{figure*}

\begin{figure*}
  \centering
  \subfigure[Drag force of simulation with speed 80 m/s]{
    \includegraphics[width=0.48\linewidth]{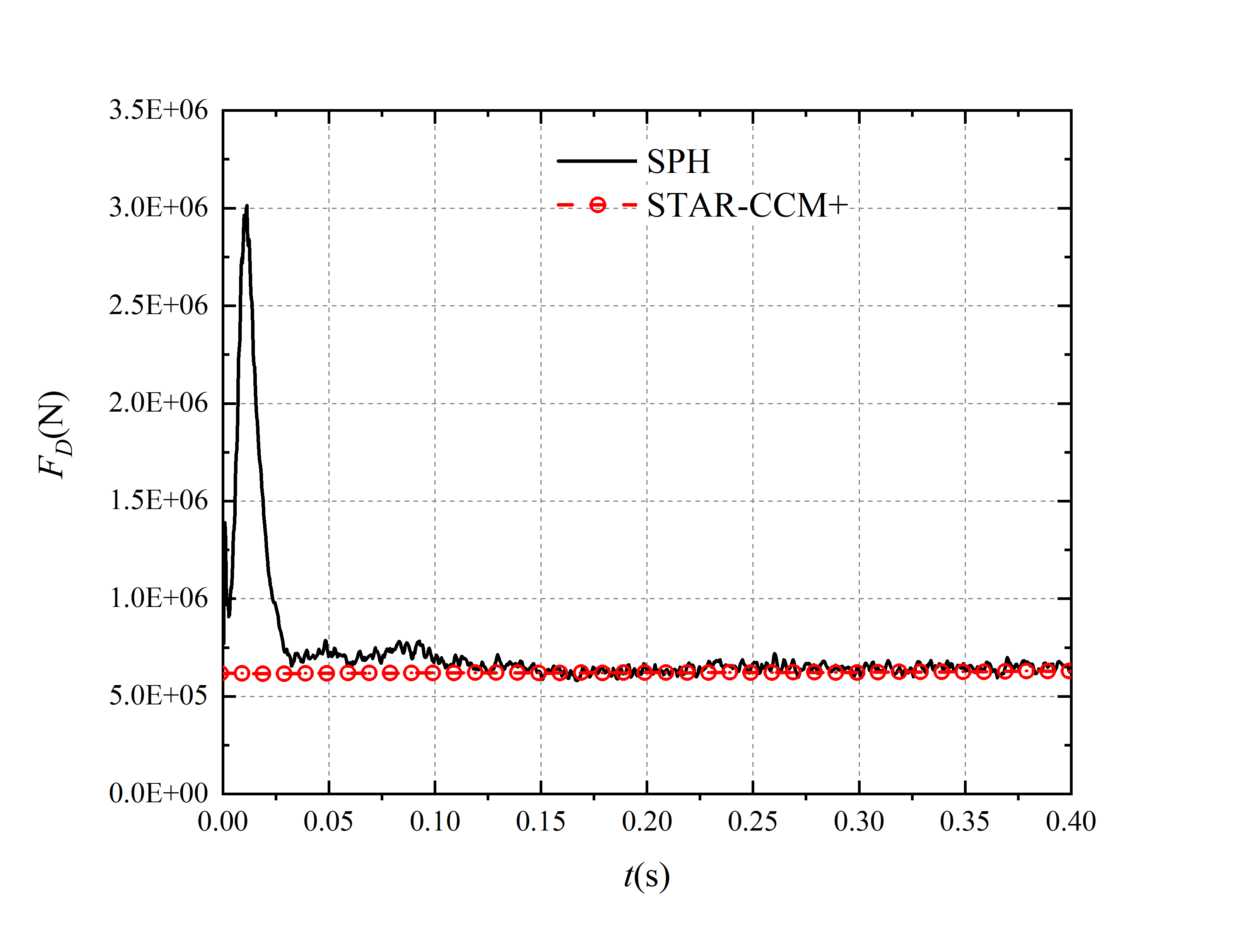}\label{fig:80mps-Drag}
  }
  \subfigure[Lift force of simulation with speed 80 m/s]{
    \includegraphics[width=0.48\linewidth]{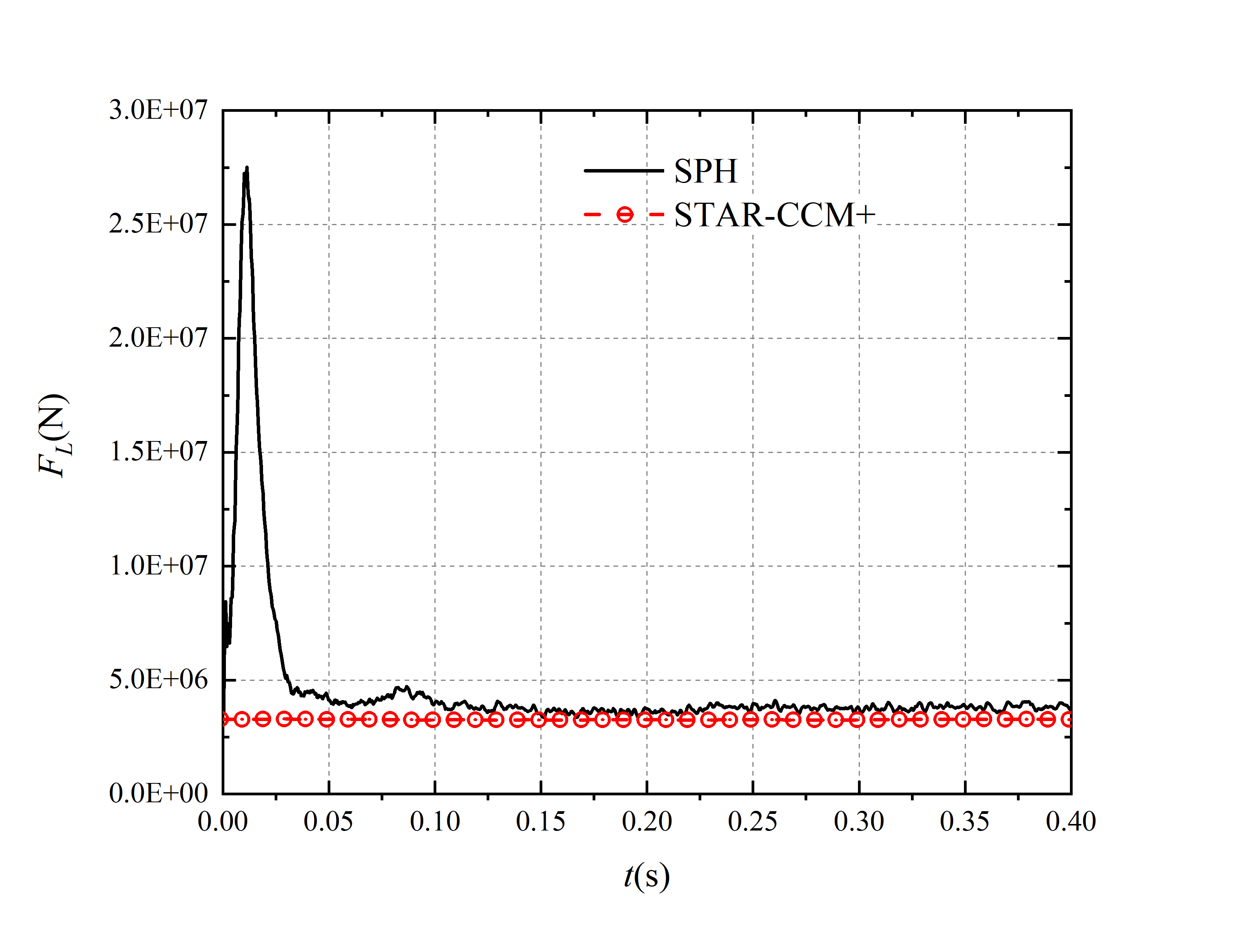}\label{fig:80mps-Lift}
  }
  \caption{The SPH results of drag and lift force validated with FVM results with a speed of 80 m/s.}
  \label{fig:80mps}
\end{figure*}l
\section{Numerical simulation and comparative analysis}

\subsection{Model parameters}
In order to reduce computing resources, parts of the seaplane in contact with water are taken into consideration and other accessories such as wings are not considered. The model is shown in Fig.\ref{fig:Seaphane-initial}.
The origin of the coordinate is located at the intersection of the longitudinal section of the seaplane and the static water surface. The $x$-axis direction points to the tail of the seaplane, and the $z$-axis direction is vertically upward. In the numerical simulation, the seaplane is fixed and the taxiing process is realized by employing uniform incoming flow.

As the world's largest seaplane in development, AG600 (shown in Fig.\ref{fig:AG600-network}) has a length of 36.9 meters and a height of 12.1 m. Its external size is comparable to that of the Boeing 737. The maximum load capacity reaches 12 tons, and the take-off speed reaches 300 km/h, which is higher than any other high-speed ship in the world.

In this paper, an in-house SPH program and commercial software STAR-CCM+ based on FVM are used to simulate different working conditions respectively, and the results of the two are compared. To compare the results more credible, two different speeds are set up, 30 m/s and 80 m/s.

The APR is utilized to improve accuracy and reduces the total calculation time and cost. A smaller particle spacing is set near the seaplane to capture the change of the free surface. A larger particle spacing is set in the part away from the seaplane to reduce the number of particles in the computational domain and improve computational efficiency.
Fig.\ref{fig:APR-Seaphane-SPH} shows the initial state of SPH simulation.
Correspondingly, the numerical tank is established and the numerical simulation of the taxiing process of the seaplane at different speeds is carried out, combined with VOF to capture the free surface.
The calculation domain setting and meshing are shown in Fig.\ref{fig:APR-Seaphane-FVM} below.
The boundary layer setting has a great impact on fluid simulation, and for large-scale high-speed taxiing, the Y+ setting of less than 100 can meet the requirements of free surface simulation
Similar to the operation in SPH simulation, the computational domain is divided into different regions, and the mesh size of the region where the free surface changes greatly are refined.

Convergence analysis is necessary before a numerical study.
In the SPH and FVM simulations, since multi-resolution techniques (APR for SPH and AMR for FVM) can be applied, the flow details can be well resolved close to the structure.  
After a convergence analysis, we adopted the particle/mesh refinement strategy as follows. 
Two refinement operations are performed in the SPH simulation. The initial particle spacing is 0.2 m, and, in the area near the seaplane, the particle spacing is refined to 0.05 m.
In the FVM simulation, considering the free surface, the mesh sizes of x, y, and z along the three directions of the coordinate axis are set to 0.0625 m, 0.0625 m, and 0.0625 m respectively.
The total number of particles required for the numerical simulation of the SPH method is around 6.85 million, and the number of grids for FVM is about 7.42 million.

\begin{figure*}
  \centering

  \subfigure[FVM results]{
    \includegraphics[width=0.98\linewidth]{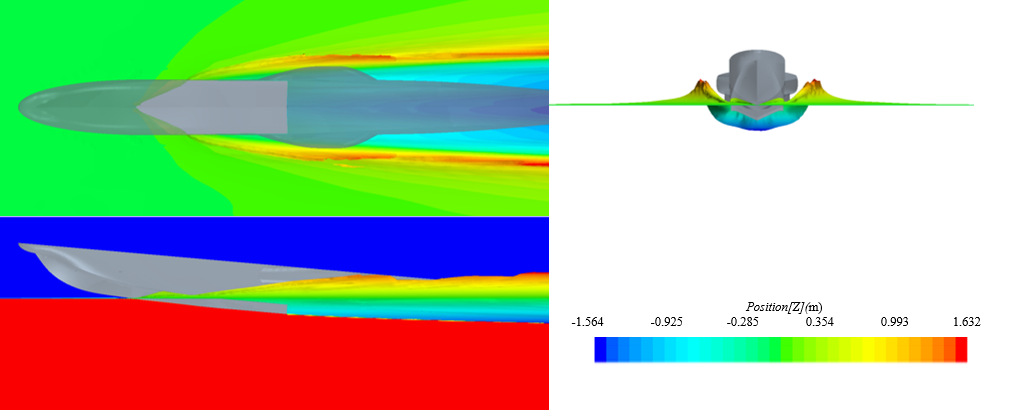}\label{FVM-results}
  }
  \subfigure[SPH results]{
    \includegraphics[width=0.98\linewidth]{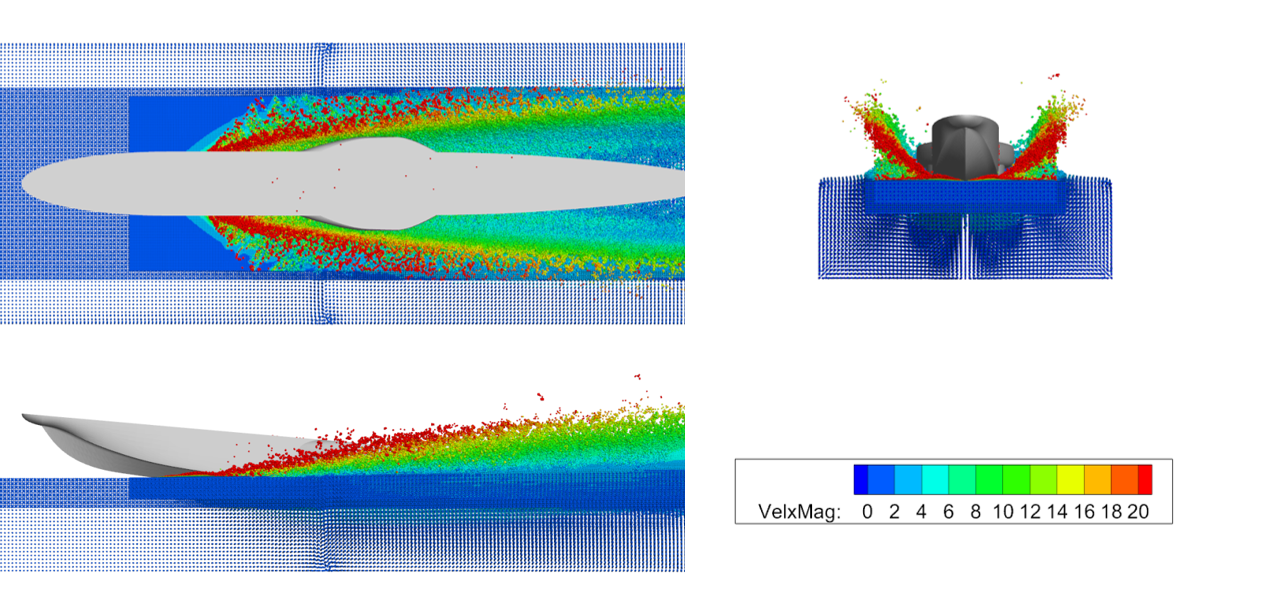}\label{fig:SPH-results}
  }
  \caption{The comparison of the splashing of free-surface between (a) SPH and (b) FVM results }
  \label{fig:splashing}
\end{figure*}

\subsection{Comparative analysis of drag and lift}
Fig.\ref{fig:30mps} and \ref{fig:80mps} show the lift and drag results obtained by the  SPH method and FVM at two different speeds. 
From a macroscopic point of view, the results of the two are very close, and the results of the SPH method reach an accuracy comparable to the commercial software. This shows that the SPH method used in this paper can be applied to calculate the lift and drag of the seaplane during the taxiing process, and the results are reliable.

In this paper, the taxiing speed of two seaplanes is numerically simulated, and the calculation results of lift and drag at different speeds are compared and analyzed.
Thanks to the special structure design of the water part of the seaplane, the lift force is greater than the drag force at high speed, whether it is 30 meters per second or 80 meters per second, which is very beneficial to the subsequent take-off process of the seaplane. As the taxiing speed increases, the lift and drag of the seaplane increase.

Analyzing the result in more detail, the resistance obtained by the SPH method will be slightly greater than the results of FVM. 
One possible reason is that the surrounding fluid splashes during high-speed taxiing, and the splashing fluid slams the seaplane.
The SPH method has a strong ability to capture these splashes, and the slamming pressure generated by the splash increases the resistance of the seaplane, which is also consistent with the results of the model test \citep{zheng_numerical_2021}. The splashing phenomenon in the taxiing process is discussed in more detail in the next section.
\subsection{Splash characteristics of flow field}
In the process of high-speed taxiing of seaplane, the splash of the free surface is inevitable.
Fig.\ref{FVM-results} shows the simulation results of FVM when the seaplane taxis at a high speed of 80 m/s. The separation of the fluid and the seaplane surface and the splash phenomenon can be found in the first part of the seaplane. The splash continues to develop to the tail of the aircraft, and finally, the seaplane forms a stable Kelvin wave on the water's surface.

Fig.\ref{fig:SPH-results} shows the splash results of three different perspectives of the seaplane at a speed of 80 meters per second calculated by the in-house SPH program. To clearly show the splash effect, the speed is processed, the parameter $VelxMax$ is set, and the calculation method is $ VelxMag = \sqrt{\left(\left(V_x -80 \right)^2 + V_y^2 + V_z^2\right)} $.
Comparing the position of the splash, it can be found that the results of SPH and FVM are consistent, both of which are generated at the head and develop to the tail.
However, there are certain differences in the wave surface and tail waveforms around the seaplane model at different speeds. For example, at high speeds (80 m/s), the width of the tail Kelvin wave is small.

\begin{figure*}
  \centering
  \includegraphics[width=0.780\linewidth]{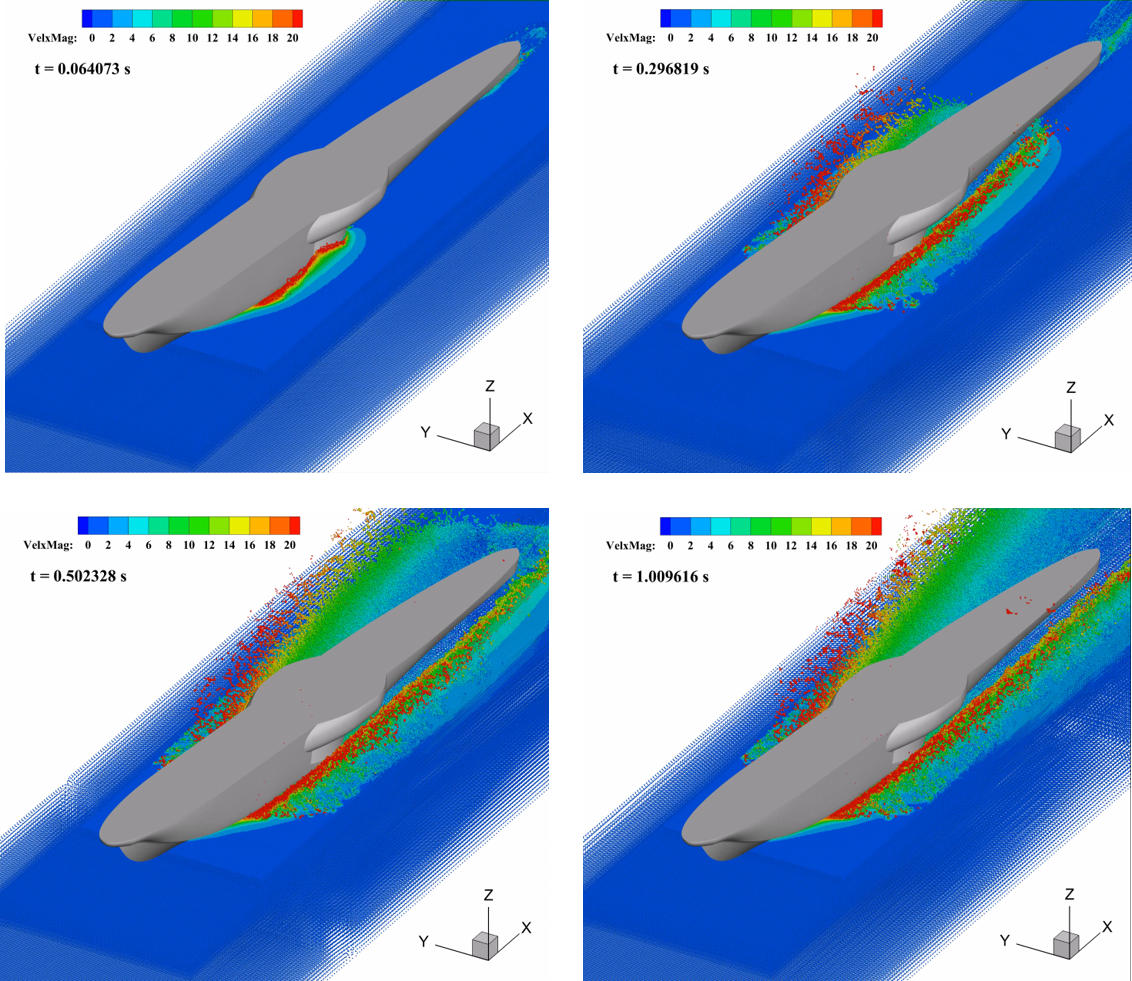}
  \caption{Splashing of the flow field at different time instants. } \label{fig:Splashing-time}
\end{figure*}

The difference is that the SPH method captures more obvious free surface splashes and flips, the width of the Kelvin wave of SPH is bigger, and the height of the splash calculated by the SPH method is significantly higher than the result of FVM. The splash state of the seaplane during high-speed taxiing calculated by the sph method is shown in Fig.\ref{fig:Splashing-time}.

Limited by the grid, it is difficult for FVM to simulate the real form of a splash when the seaplane is taxiing at high speed. 
As a meshless method of the pure Lagrangian framework, the SPH method has a natural advantage in describing violent fluid deformations. In addition, thanks to the omission of the convection term in the control equation, numerical diffusion is prevented at the interfaces when simulating high-speed flows.
For the splash phenomenon of seaplane taxiing studied in this paper, the SPH method can not only reach the same calculation accuracy comparable to the commercial software but also capture the splash phenomenon well, which is of great significance for the next step to study the influence of splash on other components of seaplane.

\section{Conclusions}
In this paper, the splash problem caused by high-speed taxiing of seaplanes is studied. SPH (particle-based method) and FVM (mesh-based method) are utilized to simulate it respectively. A numerical tank is established in FVM, and the VOF method is used to capture the change of free surface, so as to realize the numerical simulation of the splash problem.
 The free surface condition of the SPH method is automatically satisfied at the free surface, and no special treatment is needed. In order to improve the accuracy and reduce the total calculation time, Adaptive Particle Refinement is utilized. 
The drag force, lift force, and splash characteristics of a seaplane taxiing at two speeds are numerically studied.
At a larger taxiing speed, a larger splash of the free surface occurs around the seaplane. 
Compared with the FVM calculation results, the lift and drag results of SPH calculation have a small deviation within the allowable error range, which shows that the load prediction by the SPH method for the seaplane taxiing process is reliable.
Compared with the free surface splash calculated by FVM, it is shown that the SPH method can better capture the splash phenomenon during the high-speed taxiing of the seaplane, which is of great significance for the next study of the impact of a splash on other components of the seaplane.
In summary, the SPH method can obtain similar accuracy with respect to FVM in terms of the overall hydrodynamic force calculation, but the height of the splashing wave obtained by SPH is higher than that of FVM, which is closer to real physical phenomena.
Compared to FVM, the computational efficiency of SPH should be further improved. In future studies, the SPH simulation can be accelerated with GPU parallel computing. On the other hand, a coupling technique between SPH and FVM can be developed. 
The research conclusions of this paper can provide a reference and basis for the anti-spatter design of seaplanes. To a certain extent, it is expected to greatly shorten the development cycle of seaplanes and improve water taxiing performance.

\section*{Acknowledgments}

This research is supported by:
the National Natural Science Foundation of China (Grant Nos. \textcolor{blue}{52171329} and \textcolor{blue}{12002404}); 
the Natural Science Foundation of Guangdong Province of China (Grant No. \textcolor{blue}{2022A1515012084});  and the Fundamental Research Funds for the Central Universities, Sun Yat-sen University (Grant No. \textcolor{blue}{231gbi023}). 
This work was also supported by the GHfund A (Grant No. \textcolor{blue}{ghfund202302014084}). 

The OceanConnect High-Performance Computing Cluster of Sun Yat-sen University that supports scientific computing in this paper is also acknowledged.


\printcredits

\bibliographystyle{model1-num-names}

\bibliography{Xu_et_al_2023}
\end{sloppypar}
\end{document}